\documentclass[a4paper,fleqn]{cas-sc}

\usepackage[numbers]{natbib}
\usepackage{graphicx}
\usepackage{amsmath}
\usepackage{hyperref}
\usepackage{sidecap}
\usepackage{tikz}
\usetikzlibrary{calc}
\hypersetup{colorlinks=true, urlcolor=blue, citecolor=blue, linkcolor=blue}

\graphicspath{{figures/}}

\begin{document}
\let\WriteBookmarks\relax
\def\floatpagepagefraction{1}
\def\textpagefraction{.001}

\shorttitle{Wire-by-Wire CDC Tracking Efficiency}
\shortauthors{Mondal}

\title[mode=title]{Wire-by-Wire Tracking Efficiency Plots:\\
  A New Diagnostic for the Belle~II Central Drift Chamber}

\author[1]{Suryanarayan Mondal}[orcid=0000-0002-3054-8400]
\cormark[1]
\ead{suryamondal@gmail.com}
\credit{Conceptualization, Methodology, Formal analysis, Visualization, Writing -- original draft}

\affiliation[1]{organization={SSS Defence},
               city={Bengaluru},
               country={India}}

\cortext[1]{Corresponding author}

\begin{abstract}
Large detectors are often monitored at the channel level (drift time,
collected charge, and hit maps), which validates hardware but not
tracking performance.
A wire-by-wire tracking efficiency diagnostic is presented for the
Belle~II Central Drift Chamber~(CDC).
The method is directly analogous to the extrapolation-based efficiency
measurement standard in resistive-plate-chamber~(RPC) stacks developed
for the India-based Neutrino Observatory~(INO).
A reference track (helix) is extrapolated to each wire layer; the
fraction of crossings that contain an associated hit defines the
per-wire efficiency.
Implemented in the Belle~II Data Quality Monitoring~(DQM) framework and
validated on Monte Carlo simulation with controlled dead-wire conditions,
the method reveals localised tracking failures that are invisible to
conventional channel-level diagnostics.
The resulting plots provide direct feedback for run selection,
operations, and long-term ageing studies.
\end{abstract}

\begin{highlights}
\item Wire-by-wire tracking efficiency for the Belle~II CDC, computed by extrapolating helices to each layer.
\item Method adapted from the standard RPC diagnostic developed for the India-based Neutrino Observatory.
\item Reveals localised failures invisible to channel-level monitoring.
\item Integrated into Belle~II DQM; supports run selection and long-term ageing studies.
\end{highlights}


\maketitle

\section{Introduction}
\label{sec:intro}

Large detector systems are routinely treated as black boxes: if aggregate
performance metrics look reasonable, the internals are assumed to be fine.
This assumption breaks down in two related ways.
First, a localised hardware failure can be absorbed into the global
acceptance, leaving no clear signature in standard monitoring plots.
Second, in multi-detector systems one sub-detector can silently compensate for
another, creating false confidence in overall performance while leaving the root
cause undetected.

Motivation for per-component diagnostics comes from the
India-based Neutrino Observatory~(INO), whose proposed Iron Calorimeter~(ICAL)
called for approximately 28,800 glass Resistive Plate Chambers of $2\,\text{m}
\times 2\,\text{m}$ area~\cite{INOICAL}.
At a target lifetime of roughly twenty years, even a small fraction of
failed modules would carry measurable physics impact.
This motivated systematic quality assurance at every level, from
individual gas-gap integrity tests on prototypes~\cite{Mondal2019}
to layer-by-layer efficiency monitoring on the prototype stacks.

The Belle~II experiment at the SuperKEKB $e^+e^-$ collider applies the same
per-component approach.
The Central Drift Chamber~(CDC) is its primary charged-particle tracking
detector.
We show that the extrapolation-based efficiency method, standard in RPC stacks,
transfers directly to CDC wires and reveals failures invisible
to conventional channel-level monitoring.

\section{Extrapolation-Based Efficiency in RPC Stacks}
\label{sec:rpc}

\begin{figure}[h!]
  \centering
  \includegraphics[width=0.7\linewidth]{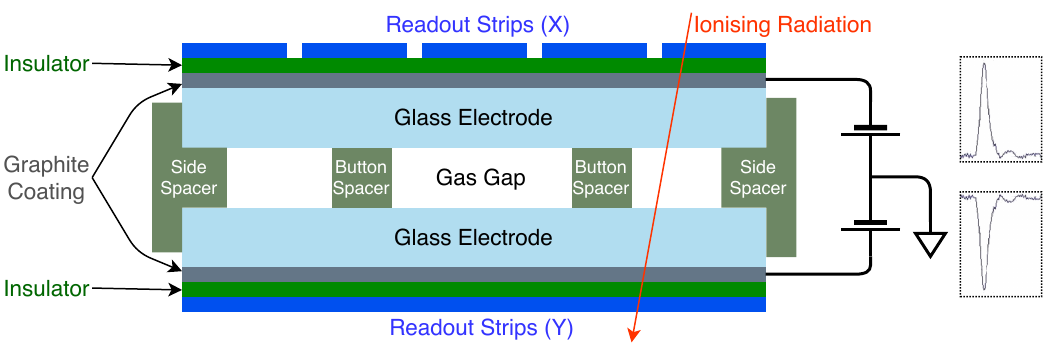}
  \caption{Schematic of a resistive plate chamber traversed by ionising
           radiation, illustrating the avalanche in the gas gap and signal
           pickup on the readout strips.}
  \label{fig:rpc-schematic}
\end{figure}

\begin{figure}[h!]
  \centering
  \includegraphics[width=\linewidth]{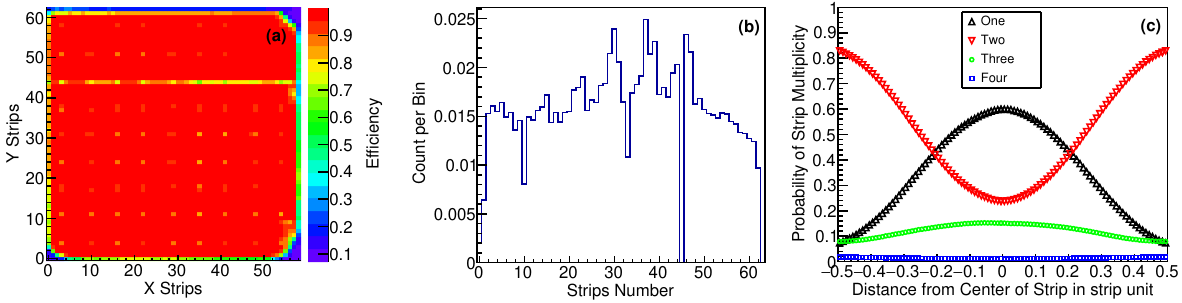}
  \caption{Per-layer efficiency (a), strip hit count (b), and
           cluster multiplicity (c) for a representative INO-ICAL
           RPC stack layer.
           Each plot answers a specific diagnostic question.}
  \label{fig:rpc-properties}
\end{figure}

An RPC detects ionising radiation through the avalanche it produces in the
gas gap between two resistive plates~(Fig.~\ref{fig:rpc-schematic}).
The standard diagnostic in a multi-layer RPC stack is extrapolation-based
efficiency measurement.
A reference track is reconstructed from all layers \emph{except} the one
under test.
That track is extrapolated to the test layer; the strip it crosses is the
\emph{expected} active element, and the layer efficiency is
$\varepsilon = N_{\text{observed}}/N_{\text{expected}}$,
where $N_{\text{observed}}$ counts crossings for which that strip recorded a hit
and $N_{\text{expected}}$ is the total number of crossings.
The efficiency, together with the strip hit counts and the cluster multiplicity
(Fig.~\ref{fig:rpc-properties}), forms a set of three plots that answer three
distinct questions:
\begin{enumerate}
  \item whether the layer detected the particle (efficiency)
  \item whether individual strips are alive (hit count map)
  \item whether the gain is at the right operating point (cluster multiplicity)
\end{enumerate}

\section{The Belle~II Central Drift Chamber}
\label{sec:cdc}

The same extrapolation principle applies to the Belle~II CDC, where
each sense wire plays the role of an RPC strip.

\subsection{Detector overview}

\begin{figure}[h!]
  \centering
  \includegraphics[width=0.48\linewidth]{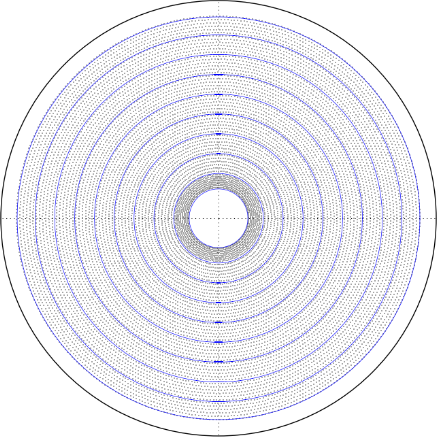}\hfill
  \includegraphics[width=0.48\linewidth]{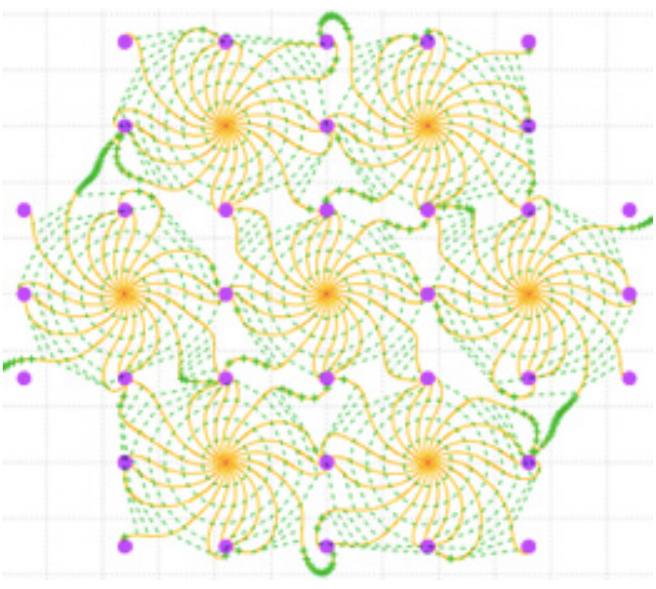}
  \caption{Left: XY cross-section of the Belle~II CDC showing the 56
           concentric wire layers grouped into superlayers, with circular
           boundaries marking the transition between axial and stereo superlayers.
           Right: drift-cell field map illustrating the hexagonal cell geometry
           and the field-wire arrangement around each sense wire.}
  \label{fig:cdc-overview}
\end{figure}

The Belle~II CDC~\cite{Belle2TDR} is a cylindrical drift chamber filling the
radial range from $\sim$16\,cm to $\sim$113\,cm.
It consists of 56 wire layers organised in 9 superlayers (one innermost axial
superlayer followed by alternating stereo and axial superlayers), providing
up to 56 hits per track.
The CDC is supplemented by the four-layer Silicon Vertex
Detector~(SVD) and the two-layer Pixel Detector~(PXD).
Together they provide the primary vertex measurement and supply
seed tracks to the CDC pattern recognition algorithms
(Fig.~\ref{fig:cdc-overview}).

\subsection{Conventional wire health monitoring}

\begin{SCfigure}[1][h!]
  \includegraphics[width=0.50\linewidth]{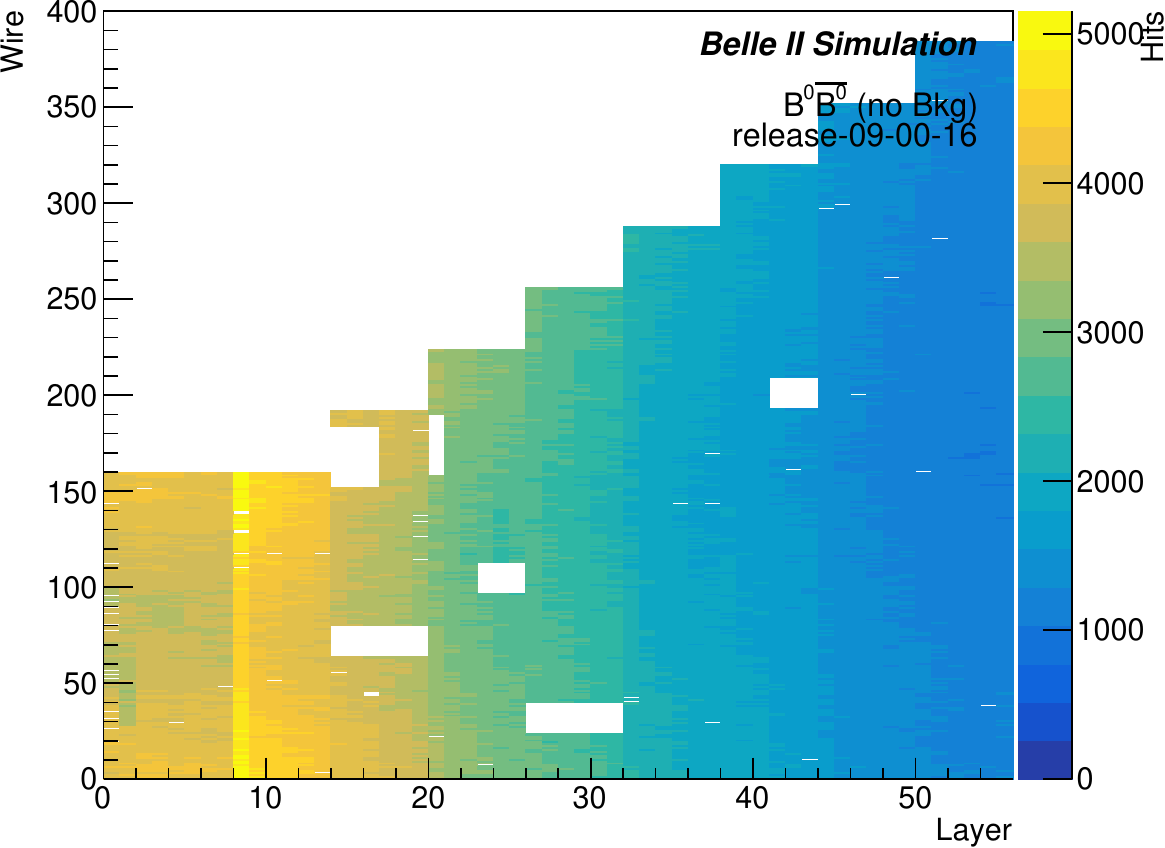}
  \caption{CDC wire hit map appearing globally healthy despite a localised
           tracking failure, illustrating why channel-level monitoring
           alone is insufficient.}
  \label{fig:cdc-hitmap}
\end{SCfigure}

Standard CDC DQM monitors each wire through:
\begin{itemize}
  \setlength{\itemsep}{0pt}
  \item \emph{Hit maps}: the number of times each wire fires per unit luminosity
        (Fig.~\ref{fig:cdc-hitmap});
  \item \emph{Drift time distributions}: sensitive to gas conditions and
        high-voltage stability;
  \item \emph{Collected charge}: sensitive to gas gain and ageing.
\end{itemize}
Each quantity confirms that the hardware is functioning within limits,
but none directly addresses whether the detector is finding the tracks it should.

\subsection{Why channel-level monitoring is insufficient}

The Belle~II tracking chain is designed for robustness.
If the CDC does not find a track, the SVD attempts independent reconstruction;
the \texttt{SVDtoCDCCKF} Combinatorial Kalman Filter~\cite{Belle2Tracking}
then tries to attach CDC hits to the SVD seed.
This resilience is a strength for physics, but a weakness for diagnostics.
A localised CDC inefficiency is partially absorbed by SVD, suppressing any
efficiency drop and leaving channel-level plots uninformative.

\section{Wire-by-Wire Efficiency Method}
\label{sec:method}

The efficiency is computed from track helices provided by the standard Belle~II reconstruction.
As a charged particle traverses a CDC wire layer, the helix is extrapolated
to identify the sense wire nearest to the crossing point.
That wire is the \emph{expected} hit: it should have recorded a signal if functioning.
If the reconstructed track has an associated hit on that wire, the crossing is
counted as \emph{observed}.
Unlike the strict RPC procedure, the layer under inspection is not excluded from
the fit: with up to 56 layers per track, the gain in fit independence is
marginal and the benefit does not justify the added complexity.
The result is accumulated in a two-dimensional histogram of
wire index vs.\ layer index (or equivalently vs.\ $\phi$), giving a
geometric view of the full CDC acceptance.
Because the efficiency is defined relative to the track, the diagnostic
needs no external reference detector or baseline run, making it suitable
for early failure detection.
The DQM module runs both online and offline, accumulating statistics
across all trigger types.

\section{Simulation Setup}
\label{sec:setup}

\subsection{Software and track selection}

Tracks are selected from the standard Belle~II reconstruction output
via the publicly available \texttt{basf2} framework~\cite{basf2}.
The results presented here use a Monte Carlo sample of
$e^+e^-\!\to B^0\bar{B}^0$ events without beam background.
Beam background is omitted because the method requires only reconstructed
tracks; its validation is independent of the underlying event conditions.

\subsection{Dead-wire conditions}

The simulation payload includes scattered genuine dead wires together with
two injected full-depth board failures, one in superlayer~2
(inner, low radius) and one in superlayer~4 (outer, high radius), each
spanning all layers of the respective superlayer.
This two-radius design quantifies how severely the track quality estimator
penalises a gap closer to the interaction point.

\section{Results and Discussion}
\label{sec:results}

\subsection{Efficiency map and localised failure}

\begin{figure}[h!]
  \centering
  \includegraphics[width=0.85\linewidth]{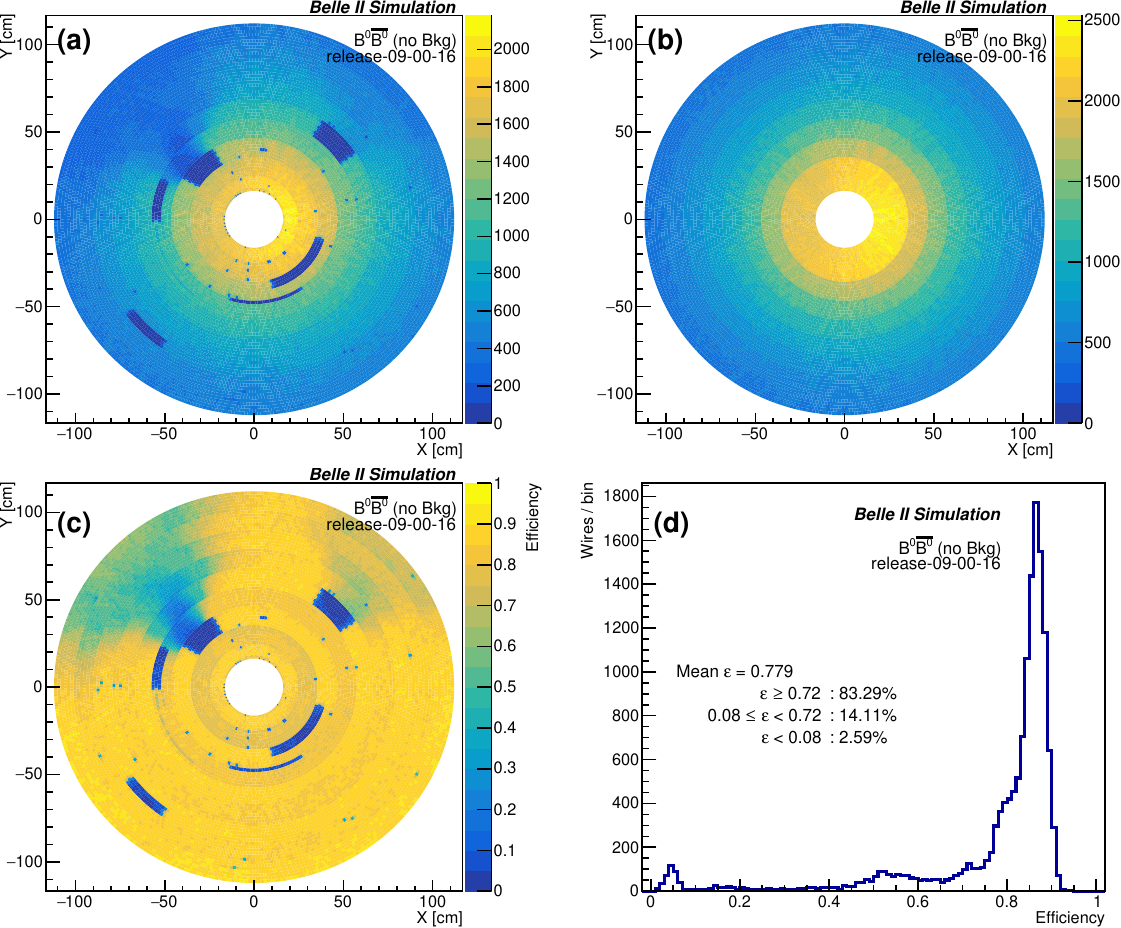}
  \caption{Wire-by-wire tracking efficiency canvas for a Monte Carlo simulation of
           $e^+e^-\!\to B^0\bar{B}^0$ events (Belle~II simulation, no beam background).
           (a) Observed wire hits.
           (b) Expected wire hits from track extrapolation.
           (c) Wire efficiency map showing clear drops at the locations of the
               two simulated dead-board regions.
           (d) One-dimensional efficiency distribution cleanly separating active
               from masked wires.}
  \label{fig:eff-map}
\end{figure}

Fig.~\ref{fig:eff-map} shows the wire-by-wire efficiency canvas for the
simulation described in Sec.~\ref{sec:setup}.
The efficiency map (c) shows drops at two azimuthal locations: a pronounced
drop near $\phi \approx +135^\circ$ from the disabled SL2 board (inner, closer
to the interaction point), and a less severe drop near $\phi \approx +42^\circ$
from the disabled SL4 board (outer).
The one-dimensional distribution (d) reveals that 83.29\% of wires have
efficiency above 0.72, while 14.11\% fall in the intermediate range
($0.08 < \varepsilon < 0.72$), and 2.59\% have $\varepsilon < 0.08$
(effectively dead); the mean wire efficiency is 77.9\%.
The 2.59\% dead-wire fraction corresponds to 371 wires out of 14\,336, consistent
with the injected dead-wire conditions.
The per-wire efficiency values allow run-by-run selection: runs where specific $\phi$ regions
fall below threshold can be excluded or down-weighted rather than rejected wholesale.
The thresholds are indicative and can be refined with operational experience.

\subsection{Investigation}

Both drops also appear in the $\phi$ distribution of the standard CDC DQM
module, but that plot alone provides no further explanation.
When a full axial superlayer is missing, CDC tracking produces incomplete
tracks; the MVA-based track quality estimator removes these as poorly formed.
SVD standalone tracking then recovers the particle, and the
\texttt{SVDtoCDCCKF} algorithm~\cite{Belle2Tracking} attaches CDC hits only up
to the gap, unable to bridge the missing layer.
Because the track does not reach the outer CDC layers, it fails to extrapolate
to the Electromagnetic Calorimeter~(ECL), preventing calorimeter association;
this is particularly serious for electron and photon-related analyses.
The efficiency map alone is therefore sufficient to localise the failure and
reveal its downstream consequences.
This makes it a direct input to operational decisions.

\subsection{Operational impact}

The efficiency histograms are now part of the Belle~II CDC DQM.
They provide three classes of information:
\begin{itemize}
  \setlength{\itemsep}{0pt}
  \item \textbf{Immediate failure detection}: a disabled board or dead high-voltage
        channel produces a drop over a contiguous region of wires, visible
        within a single run.
  \item \textbf{Run selection}: analysts can exclude or down-weight runs where
        specific $\phi$ regions show reduced efficiency, rather than applying
        a global run quality flag.
  \item \textbf{Ageing and long-term studies}: the per-wire efficiency trend
        accumulated over many runs provides a sensitive probe of wire ageing,
        gas gain evolution, and detector degradation.
\end{itemize}

\section{Summary}
\label{sec:summary}

The extrapolation-based efficiency method, established in the INO RPC
prototype programme~\cite{INOICAL,Mondal2019}, transfers directly to
drift-wire detectors.
Applied wire by wire to the Belle~II CDC, it exposes localised hardware
failures invisible to hit-map monitoring, including their downstream effect
on track quality rejection and calorimeter association.
The method is implemented in the Belle~II DQM framework~\cite{basf2}
and provides immediate, wire-resolved feedback for operations and
physics analysis.
The approach generalises to any tracking detector with individually
instrumented channels, enabling reconstruction-level quality monitoring
beyond channel-level plots.
The failure mode also motivates retraining the MVA-based CDC track
quality estimator to handle partially missing axial superlayers, improving
track retention under realistic detector conditions.

\printcredits

\bibliographystyle{cas-model2-names}
\bibliography{refs}

\end{document}